\documentclass[aps,prb,10pt, reprint, floatfix,preprintnumbers,showpacs, 
amsfonts, amsmath,amssymb, citeautoscript, superscriptaddress]{revtex4-1}
\usepackage{graphicx}
\usepackage[utf8x]{inputenc}
\usepackage{color}
\usepackage{etoolbox}
\usepackage{multirow}
\usepackage{braket}
\usepackage{soul}
\apptocmd{\sloppy}{\hbadness 10000\relax}{}{}

\newcommand*{\bo}{\mathbf}
\newcommand*{\bs}{\boldsymbol}

\newcommand*{\ua}{\uparrow}
\newcommand*{\da}{\downarrow}

\newcommand*{\meff}{m_{\text{eff}}}
\newcommand*{\Kp}{\widetilde{\mathbf{K}}_1}
\newcommand*{\Km}{\widetilde{\mathbf{K}}_{-1}}
\newcommand*{\Kt}{\widetilde{\mathbf{K}}_{\tau}}
\DeclareMathOperator{\Tr}{Tr}
\DeclareMathOperator{\I}{Im}
\DeclareMathOperator{\R}{Re}
\DeclareMathOperator{\sgn}{sgn}

\begin{document}

\title{RKKY interaction and intervalley processes in p-doped transition 
metal dichalcogenides}
\date{\today}

\author{Diego Mastrogiuseppe}
\author{Nancy Sandler}
\author{Sergio E.\ Ulloa}
\affiliation{Department of Physics and Astronomy, and Nanoscale
and Quantum Phenomena Institute, \\ Ohio University, Athens, Ohio
45701--2979}

\begin{abstract}
We study the Ruderman-Kittel-Kasuya-Yosida (RKKY) interaction in p-doped 
transition metal dichalcogenides such as MoS$_2$ and WS$_2$. We consider 
magnetic impurities hybridized to the  Mo $d$-orbitals characteristic of the 
valence bands. Using the Matsubara Green's function formalism, we obtain the 
two-impurity interaction vs their separation and chemical potential of the 
system, accounting for the important angular dependence which reflects the 
underlying triangular lattice symmetry. 
The inclusion of the valence band valley at the $\Gamma$ point results in a 
strong enhancement of the interaction.
Electron scattering processes transferring momentum between valleys at 
different symmetry points give rise to complex spatial oscillation patterns. 
Variable doping would allow the exploration of rather interesting
behavior in the interaction of magnetic impurities on the surfaces of these materials, 
including the control of the interaction symmetry, 
which can be directly probed in STM experiments.
\end{abstract}  

\pacs{75.30.Hx, 75.20.Hr, 75.75.-c, 75.70.Tj}

\maketitle

\noindent \emph{Introduction}.---The Ruderman-Kasuya-Kittel-Yosida (RKKY) interaction 
\cite{Ruderman1954,Kasuya1956,Yosida1957}, or indirect exchange, 
describes the effective coupling of two magnetic moments mediated by 
conduction electrons in a metal. Under certain conditions, this 
interaction can give rise to effects such as itinerant magnetic 
order, and giant magnetoresistance \cite{Grunberg1986, Baibich1988, Bruno1991}, 
with important technological applications.
As such, it directly impacts the field of spintronics \cite{Wolf2006}, 
allowing information transfer between spins in a 
controlled manner.

The RKKY interaction depends on the dimensionality and underlying band structure
of the host material. For example, in conventional two dimensional metals, 
it oscillates 
with inter-impurity separation $r$ with a characteristic wavelength ($\approx \lambda_F/2$, half the 
Fermi wavelength in the host). The oscillation expresses the alternation
between ferromagnetic (FM) and antiferromagnetic (AFM) coupling, decreasing
as $r^{-2}$ \cite{Fischer1975}. Remarkably, complex band structures can give rise 
to nonstandard behavior. In graphene, for instance, the RKKY interaction decays as
$r^{-3}$ for the charge neutral system, while more conventional behavior appears in the 
doped or gapped cases \cite{Dugaev2006, Saremi2007, Black-Schaffer2010, 
Uchoa2011, Sherafati2011a, Sherafati2011, Power2012, Roslyak2013, Kogan2013, 
Power2013}. 

Other newly isolated two-dimensional layered crystals \cite{Novoselov2005} allow one
to explore even more interesting scenarios. 
A prominent example is given by transition metal 
dichalcogenides (TMDs), a family of materials where the combination of 
hybridization and strong spin-orbit interaction, due to the heavy 
transition metals atoms, results in a band structure 
with strong coupling of spin and valley degrees of freedom \cite{Xiao2012}.
The RKKY interaction in TMDs has been recently characterized in particular for MoS$_2$. 
\cite{Parhizgar2013, Hatami2014}
Parhizgar {\em et al}.\ report that the spin-spin interaction can be seen to
include three different terms: Ising, XY and Dzyaloshinskii-Moriya 
components \cite{Parhizgar2013}, all found to decay as $r^{-2}$.
In contrast, Hatami {\em et al}.\ finds that, while the out-of-plane component 
decays as $r^{-2}$, the in-plane interaction decays as $r^{-5/2}$, a 
disagreement perhaps produced by their disregard of intervalley scattering 
\cite{Hatami2014}. 

These discrepancies reveal the subtleties involved in properly accounting for all 
relevant scattering processes 
that determine the final magnetic arrangement. Interestingly, 
processes that consider the valence band valley centered at the $\Gamma$ point, 
especially important when 
considering the p-doped case, have been neglected in previous studies. 
The $\Gamma$ valley is known to lie not far removed in energy from the valleys 
at the Brillouin zone corners in MoS$_2$ and  WS$_2$  
\cite{Cheiwchanchamnangij2012,Yun2012, Shi2013, Kormanyos2013, Zahid2013}.
This valley plays a star role in the transition to the indirect gap 
behavior in bi- and multi-layers of these materials.

We analyze the RKKY interaction for p-doped TMDs 
\cite{Laskar2014,Ye2012,Zhang2013,Braga2012,Sik2012,Jo2014}, and
focus on the case of MoS$_2$ for which the relevant structure parameters are 
well known. The unavoidable contribution of the $\Gamma$ valley significantly increases 
the overall interaction strength when the Fermi level 
is set to populate this valley. Moreover, it provides 
extra channels for electron scattering processes, giving rise to complex 
spatial and energy modulation patterns for the anisotropic exchange coupling constants.
Remarkably, the inclusion of this valley allows for the possibility of 
isotropic and in-plane magnetic order, not possible in its absence.
These behaviors are easily tunable by sweeping the Fermi level and turn out to be 
important for even relatively low p-doping levels.

\begin{figure}[hbt]
\includegraphics[width=0.3\textwidth]{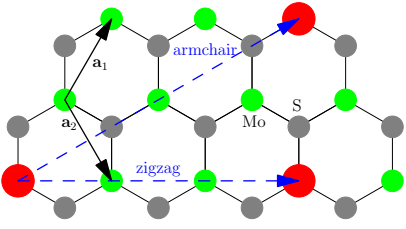}
\caption{(Color online) Magnetic impurities (red circles) hybridized to Mo 
$d$-orbitals. Blue dashed arrows show two high-symmetry directions, zigzag and 
armchair, along which we compare the effective interaction between local 
moments. Black solid arrows indicate unit vectors.}
\label{fig:imp}
\end{figure}

\noindent
\emph{Theoretical description.}---%
The basic structure of TMDs in their 2D form (elemental `monolayer') is a 
triangular layer of transition metal atoms sandwiched between two triangular 
layers of chalcogen atoms (see Fig.\ \ref{fig:imp}).
The first Brillouin zone for the monolayer crystal is hexagonal \cite{Mattheiss} 
with two nonequivalent $K_1$ and $K_{-1}$ 
valleys, in which most of the low energy physics takes place. Lack of reflection
symmetry along the $z-$axis in the unit cell produces a splitting of the metal 
$d$-orbitals resulting in a direct gap at $K_1$ and $K_{-1}$ 
valleys. The high atomic number of the transition metal produces
a sizable spin-orbit coupling which further splits the valence bands 
into two with opposite spin projection \cite{Cheiwchanchamnangij2012}. 
These two effects result in a strong spin-valley 
coupling, while spin remains a good quantum number \cite{Xiao2012}.

Several recent {\em ab initio} calculations show that the (spin-degenerate)
valence band valley at the $\Gamma$ point, 
also contributes to the low energy physics 
\cite{Cheiwchanchamnangij2012,Yun2012, Shi2013, Kormanyos2013, Zahid2013}.
The $\Gamma$ valley participates in virtual transitions even at low p-doping 
levels (or gating ranges) common in experiments 
\cite{Laskar2014,Ye2012,Zhang2013,Braga2012,Sik2012,Jo2014}.

\begin{figure}[htb]
\includegraphics[width=0.48\textwidth]{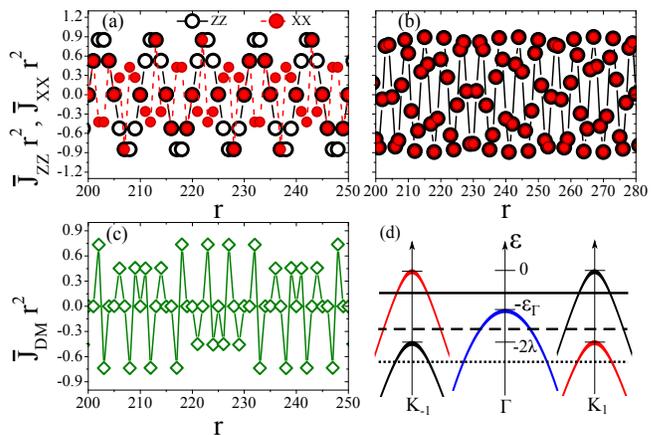}
\caption{(Color online) ZZ and XX components of the RKKY interaction as 
function of impurity separation $r$, along (a) zigzag and (b) armchair 
directions.  
(c) DM component in the zigzag direction. The latter vanishes in the armchair 
direction. In all cases the interaction amplitude decays as $r^{-2}$. The Fermi 
level $\epsilon_F \simeq -0.067$, crosses the uppermost $K_\tau$ valleys, 
without intersecting the valley at the $\Gamma$ point, as indicated by the solid 
line in (d). (d) Schematic low energy band structure for MoS$_2$ and WS$_2$, 
showing the spin inversion of the valence bands at $K_1$ and $K_{-1}$ valleys. 
The black (red) curve corresponds to up (down) spin projection. The blue valley 
at $\Gamma$ is quadratic and spin degenerate. Dashed and dotted lines indicate 
higher p-doping levels discussed in Fig.\ \ref{fig:EF_m0d16_vs_R_total_zigzag} 
and below.}
\label{fig:EF_m0d06248_gp1sq_vs_R}
\end{figure}

The proposed effective low energy Hamiltonian to describe these properties is given by:
\begin{equation}
H_0 = \sum_{q,\tau} \psi^\dagger h_{K_\tau}(q) \psi + \sum_k \phi^\dagger 
h_\Gamma(k) \phi,
\end{equation}
where
\begin{equation}
h_{K_\tau}(q) = 
\left(\!\!
  \begin{array}{cccc}
  \xi & a \tau q e^{-i\tau \theta} & 0 & 0\\
  a \tau q e^{i\tau \theta} & \lambda(\tau-1) & 0 & 0\\
  0 & 0 & \xi & a \tau q e^{-i\tau \theta}\\
  0 & 0 & a \tau q e^{i\tau \theta} & -\lambda(\tau+1)
  \end{array}
\!\!\right),
\end{equation}
is the matrix near the $K_\tau$ valleys, $\tau=\pm 1$ is the valley index; 
 $q=|\bf q|$, is the modulus of the reduced wave vector measured from $K_\tau$, 
and $\theta = \arctan (q_y/q_x)$. 
The spinor bases are arranged as $\psi = (z^2\!\ua, xy\!\ua, z^2\!\da, 
xy\!\da)^T$, where $z^2$ ($xy$) stands for $\Ket{d_{3z^2-r^2}}$ 
($\left[\Ket{d_{x^2-y^2}}+ i\tau \Ket{d_{xy}}\right]/\sqrt{2}$) Mo $3d$ 
orbitals, and $\phi = (p_{xy}\!\ua, d_{z^2}\!\ua, p_{xy}\!\da, d_{z^2}\!\da)^T$, 
where $p_{xy}$ are S $p_x,p_y$  orbitals. The up/down arrows indicate the 
$z$-spin projection.  Energies are expressed throughout in units of the 
nearest-neighbor hopping amplitude $t$, $a$ is the nearest Mo-Mo distance, 
$\xi=\Delta - \lambda$, where $\lambda$ is the spin-orbit coupling constant, and 
$\Delta$ stands for the gap. Typical values for MoS$_2$ are $a\simeq 3.2 \text{ 
\AA}$, $t\simeq 1.1 \text{ eV}$, so that 
$\Delta\simeq 1.5$, and $\lambda \simeq 0.07$.
The energies have been shifted such that the top of the valence bands at the 
$K_\tau$ points lie at zero energy.
At the $\Gamma$ point we have \cite{Kormanyos2013}
\begin{equation}
h_{\Gamma}(k) = E_\Gamma(k)
\left(\!\!
  \begin{array}{cccc}
  0 & 0 & 0 & 0\\
  0 & 1 & 0 & 0\\
  0 & 0 & 0 & 0\\
  0 & 0 & 0 & 1
  \end{array}
\!\!\right),
\end{equation}
where $k$ is the modulus of the wave vector measured from the $\Gamma$ 
point, $E_\Gamma(k) = \hbar^2 k^2/(2t \meff)+\epsilon_\Gamma$. $\meff$ 
is the (negative) effective mass, and $\epsilon_\Gamma$ sets the relative
position of the $\Gamma$ and $K_\tau$ valleys ($\epsilon_\Gamma \approx 0.1$ in 
MoS$_2$). The conduction matrix elements were discarded due to the large gap 
between conduction and valence bands. A schematic representation of the valence 
band structure around the three relevant points in the Brillouin zone is shown 
in Fig.\ \ref{fig:EF_m0d06248_gp1sq_vs_R}(d).

Next, we consider two spin-$1/2$ s-wave magnetic impurities hybridized to Mo 
atoms, given 
that relevant Bloch states at low energies are composed mainly from 
admixtures of $d$ orbitals from these atoms. We choose two high symmetry directions 
connecting these local moments, zigzag and armchair, 
to show characteristic results, although many other directions are clearly possible---
see Fig.\ \ref{fig:imp}.
The interaction between each  magnetic atom and conduction electron 
spins in the host is described by a contact interaction 
$H_{\text{int}} = J \sum_{j=1,2} \bo S_j \cdot \bo s(\bo R_j)$,
where $\bo s(\bo r) = \frac{1}{2} \sum_i \delta(\bo r - \bo r_i)\bs \sigma_i$
represents the spin density for electron $i$ ($\hbar=1$), and $\bo S_j$ is the 
localized spin at site $\bo R_j$. For simplicity, we assume the same 
exchange coupling $J$ for valence electrons on both $d_{xy}$ and $d_{z^2}$ Mo 
orbitals.
One can treat $H_{\text{int}}$ as a perturbation of $H_0$; obtaining at second
order an effective interaction between the localized spins \cite{Mattis}
\begin{equation}
 H_{\text{RKKY}} = J^2 \sum_{\alpha,\beta} S_1^{\alpha}\, 
\chi_{\alpha,\beta}(\bo R)\, S_2^{\beta},
\end{equation}
where $\chi_{\alpha,\beta}$ is the static spin susceptibility tensor of the
electron gas, with $\alpha,\beta$ representing the Cartesian 
components, and $\bo R$ is the vector connecting the magnetic moments.
The susceptibility can be calculated from the unperturbed real space 
retarded Green's function  \cite{Imamura2004, Parhizgar2013}
\begin{equation}
\begin{split}
 &\chi_{\alpha,\beta}(\bo R) = \\
 &-\frac{1}{\pi} \Tr \left[ \int_{-\infty}^{\epsilon_F} \!\!d\epsilon\:
 \I\left\{\sigma_\alpha  G(\bo R,\epsilon^+) \sigma_\beta G(-\bo R,\epsilon^+) 
\right\}\right],
\end{split}
\end{equation}
where $\epsilon^+=\epsilon+i 0^+$, and $\sigma$ are Pauli
matrices for the spin degree of freedom. $G$ stands 
for the $2\times 2$ Green's function matrix for the valence 
sector---processes that involve the conduction band are ignored, as they are 
strongly suppressed by the substantial energy gap. 
Different components of the susceptibility are
$\chi_{\alpha,\beta}(\bo R) =  -\frac{1}{\pi} \int_{-\infty}^{\epsilon_F} 
\!\!d\epsilon\:  \I A_{\alpha,\beta}(\bo R,\epsilon^+),$
with
\begin{align}\label{eq:Azz}
  A_{z,z} &= \sum_s G_s(\bo R,\epsilon^+) G_s(-\bo R,\epsilon^+),\\ 
 A_{x,x} &= A_{y,y} = \sum_s G_s(\bo R,\epsilon^+) G_{-s}(-\bo 
R,\epsilon^+),\\
 A_{x,y} &= -A_{y,x} =  -i\sum_s s\: G_s(\bo R,\epsilon^+) G_{-s}(-\bo 
R,\epsilon^+) ,
\end{align}
where $G_s(\bo R,\epsilon^+) =  G_\Gamma(\bo R,\epsilon^+) + \sum_\tau 
G_{\tau,s}(\bo R,\epsilon^+)$, and $s =\ua, \da$.
The effective anisotropic spin interaction between localized
moments includes Ising (ZZ), XX and Dzyaloshinskii-Moriya (DM) interactions, such 
that the RKKY Hamiltonian can be expressed as \cite{Parhizgar2013}
\begin{equation}
\begin{split}
 H_{RKKY} =& J_{XX} (S_1^x S_2^x + S_1^y S_2^y) + J_{ZZ} S_1^z S_2^z \\
 &+ J_{DM} (\bo S_1 \times \bo S_2)_z,
\end{split}
\end{equation}
where $J_{XX} = J^2 \chi_{x,x}$, $J_{ZZ}= J^2 \chi_{z,z}$, and $J_{DM} = J^2 
\chi_{x,y}$. Notice that the XX and DM terms compete as to favor  
(anti)parallel or perpendicular alignment of the spins respectively in the $xy$ 
plane at different impurity separations $\bo R$, creating in general an in-plane twisted 
spin structure, depending on their relative strength and sign.

It is convenient to obtain the Green's functions in momentum space and then 
Fourier-transform back to real space. \cite{suppl}
There are only two independent Green's functions at $K_1$ and
$K_{-1}$, $g_{-1,-s}(\bo R,\epsilon^+) = g_{1,s}(\bo R,\epsilon^+)$.
Omitting the energy variable for convenience, one obtains
$G_s(\bo R) = G_\Gamma(R) + \sum_{\tau} e^{i\bo K_\tau \cdot \bo R}  
g_{\tau,s}(R)$, 
and using Eq.\ \eqref{eq:Azz}, we arrive at
\begin{equation}\label{eq:Azz_expanded}
\begin{split}
\I\: &A_{z,z}= 2\Bigl(I_{G_\Gamma;G_\Gamma} + \left[\cos(\bo K_1 \cdot 
\bo R) + \cos(\bo K_{-1} \cdot \bo R)\right]\\
 &\times \left(I_{G_\Gamma; g_{1,\ua}} + I_{G_\Gamma; g_{-1,\ua}}\right) + 
I_{g_{1,\ua};g_{1,\ua}} + I_{g_{-1,\ua};g_{-1,\ua}}\\
&+ 2\cos\left[(\bo K_{1}-\bo K_{-1}) \cdot \bo R\right] I_{g_{1,\ua}; 
g_{-1,\ua}}\Bigr),
\end{split}
\end{equation}
where we have defined $I_{u; v}(\bo R,\epsilon) \equiv \I [u(\bo R,\epsilon) 
v(\bo R,\epsilon)]$ with $u,v = \{G_\Gamma; g_{1,\ua}; g_{-1,\ua}\}$.
A similar procedure yields the $A_{x,x}$ and $A_{x,y}$ 
components. The cosines are angular coefficients that modulate the 
integral kernels $I_{u; v}$, depending on the relative direction of the 
impurities. An interesting feature of these expressions is that the underlying 
axial symmetries eliminate the DM (or XY) components for impurities arranged 
along armchair directions \cite{suppl}.

\noindent
\emph{Fixed Fermi level.}---%
We define the dimensionless exchange interactions as $\bar J_i 
=-\frac{\Omega^2}{4\pi^3 J^2} J_i$, where $i= (ZZ,XX,DM)$, and $\Omega$ is 
the area of the first Brillouin zone.
Let us first analyze the case in which the Fermi level does not intersect the 
$\Gamma$ valley,  i.e.\ with $-\epsilon_\Gamma<\epsilon_F<0$, as indicated by 
the solid horizontal line in Fig.\ \ref{fig:EF_m0d06248_gp1sq_vs_R}(d). 
$I_{g{1,\ua},g{1,\ua}}$ is the only kernel contributing to the interaction.
Figures \ref{fig:EF_m0d06248_gp1sq_vs_R}(a) and (b) show the
ZZ and XX components of the RKKY interaction vs impurity separation along the 
zigzag and armchair directions respectively. The Fermi level is fixed at 
$\epsilon_F \simeq -0.067$, and $\bar J_i r^2$ is plotted as a function of the 
dimensionless distance $r$ ($=R/a$), for large separations. The nearly constant 
amplitude reflects that the interaction decays as $1/r^2$. In the zigzag case, 
the XX angular coefficients are related by the sequence $\{1,-1/2,-1/2,\cdots\}$ 
with the ZZ ones (which are constant) \cite{suppl}, so that the ZZ component 
tend to dominate over the XX. 
In the armchair direction, both ZZ and XX components coincide. Moreover, on 
sites in which $\int_{-\infty}^{\epsilon_F} d\epsilon\: 
I_{g_{1,\ua};g_{1,\ua}}(r,\epsilon)$ vanishes, both the ZZ and XX components 
vanish.
Figure \ref{fig:EF_m0d06248_gp1sq_vs_R}(c) shows the DM component in the zigzag 
direction, with a sequence $\{0,1,-1.\cdots\}$ with respect to ZZ. As mentioned, 
the symmetry of the lattice forces this component to vanish along the armchair 
direction.

In order to examine the spatial oscillations, it is convenient to define 
$q^F_{\pm 1} \equiv q_{\pm}(\epsilon_F)$ as the Fermi wave vector for the 
valleys with quantum numbers $\tau=\pm 1, s=\ua$, and $\tau=\mp 1, s=\da$, and 
$k_\Gamma^F \equiv k_\Gamma(\epsilon_F)$, the Fermi wave vector for the $\Gamma$ 
valley.
With $\epsilon_F = -0.067$, the modulation wavelength is $\Lambda \simeq 
10$ in the zigzag direction, as observed in Fig.\ 
\ref{fig:EF_m0d06248_gp1sq_vs_R}(a) and (c), and consistent with $\Lambda = 
\pi/q_1^F$. The modulation can be described by a sinusoidal function 
$\int_{-\infty}^{\epsilon_F} d\epsilon\: I_{g_{1,\ua};g_{1,\ua}}(r,\epsilon) 
\simeq c_1 r^{-2} \sin[2q_1^F r] = c_1 r^{-2}\sin[2\pi 
r/\Lambda].$ 
The amplitude here, $c_1\simeq 0.45$, is nearly independent of 
the Fermi energy. 
Along the armchair direction the modulation of the interimpurity interaction 
exhibits a more complex pattern,
as observed in Fig.\ \ref{fig:EF_m0d06248_gp1sq_vs_R}(b).  
Going from the zigzag to armchair directions amounts 
to replacing $r$ by $\sqrt{3}r$, which can be seen as a shift of 
$q_1^F$ to $\sqrt{3}q_1^F$ in the argument of the integral kernels \cite{suppl}, 
giving an effective $k_{F}$ that is larger (and incommensurate) than in the 
zigzag case. The incommensurate value also introduces aliasing effects.

\begin{figure}[htb]
\includegraphics[width=0.45\textwidth]{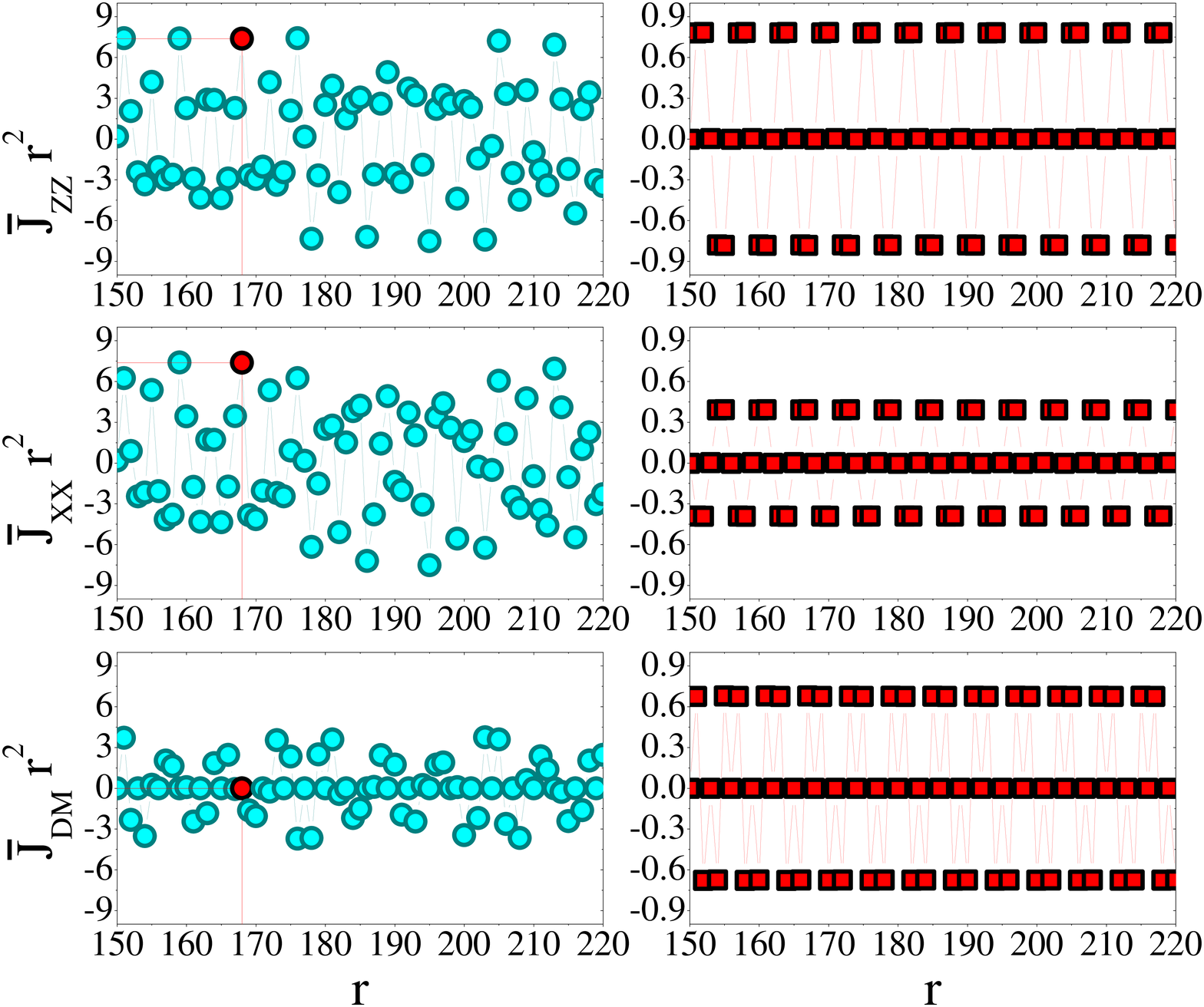}
\caption{(Color online) ZZ, XX, and DM components of the RKKY interaction, as a 
function of separation in the zigzag direction.  
$\epsilon_F = -0.174$, as indicated by dashed line in Fig.\ 
\ref{fig:EF_m0d06248_gp1sq_vs_R}(d). Left panels show the full interactions,  
including contributions of the $\Gamma$ valley. The red horizontal and vertical 
lines indicate a fully isotropic interaction point. Right 
panels show the same quantities without including the $\Gamma$ valley. Notice 
the different vertical scales.}
\label{fig:EF_m0d16_vs_R_total_zigzag}
\end{figure}

Fig.\ \ref{fig:EF_m0d16_vs_R_total_zigzag} shows results at $\epsilon_F \simeq 
-0.174$, such that the Fermi level intersects the band at the 
$\Gamma$ point [dashed line in Fig.\ \ref{fig:EF_m0d06248_gp1sq_vs_R}(d)], for 
impurities aligned along the zigzag direction. 
The right panels show the $r$ dependence of the different interaction 
components, \emph{without} the contribution of the $\Gamma$ valley, while the 
left panels show the full interaction.  The inclusion of the $\Gamma$ valley not 
only increases significantly ($\times 10$) the amplitude of the modulation 
for all the interactions, but also produces a rather complex oscillatory 
pattern, due to the additional electron scattering processes between states at 
$\Gamma$ and $K_\tau$ points. 
The integral kernels contributing significantly in this regime are 
$I_{G_\Gamma;G_\Gamma}$, $I_{G_\Gamma;g_{1,\ua}}$, and 
$I_{g_{1,\ua};g_{1,\ua}}$ \cite{suppl}.
A sinusoidal fit gives
$\int_{-\infty}^{\epsilon_F} d\epsilon\: I_{G_\Gamma;g_{1,\ua}} \simeq c_2 
r^{-2} \sin\left(\left[q_1^F+k^F_\Gamma \right] 
r\right)$,
with a wavelength given by $\Lambda=2\pi/[q_1^F+k_\Gamma^F ]\simeq 4.92$, 
and $c_2\simeq 0.24$; 
$c_2$ is found to be strongly dependent on the Fermi energy.
In the limit $\epsilon_F \rightarrow -\epsilon_\Gamma$, 
the $\Gamma$ to $K_\tau$ scattering processes
produce an unusual spatial decay $r^{-5/2}$. 
However, the weight of this component is small compared to the ones in which the 
electronic processes take place within the same band valley, so that the expected 
$r^{-2}$ decay dominates. 
Notice that the inclusion of the scattering processes at $\Gamma$ allows for 
special impurity separations in which the DM term vanishes, and $J_{XX} = 
J_{ZZ} = J_{YY}$, rendering a fully isotropic exchange interaction between them 
(see for example $r = 168$ in the figure). This feature is a consequence of the 
spin degeneracy at this valley that effectively cancels the DM component.
Similar features are observed for impurities separated along the armchair direction.

At higher p-doping, $\epsilon_F<-2\lambda$ [dotted line in Fig.\ 
\ref{fig:EF_m0d06248_gp1sq_vs_R}(d)], all valleys contribute to the indirect 
exchange, and the interaction exhibits very complex modulation patterns. The 
oscillations are dominated by
$\int_{-\infty}^{\epsilon_F} d\epsilon\: 
I_{\Gamma;g_{-1,\ua}}(r,\epsilon)\simeq c_3 r^{-2} 
\sin\left(\left[q_{-1}^F+k^F_\Gamma \right] 
r\right)$, $\int_{-\infty}^{\epsilon_F} d\epsilon\:
I_{g_{1,\ua};g_{-1,\ua}}(r,\epsilon)\simeq  c_4 r^{-2}
\sin\left(\left[q_1^F+q_{-1}^F\right] r\right)$,
and $\int_{-\infty}^{\epsilon_F} d\epsilon\: 
I_{g_{-1,\ua};g_{-1,\ua}}(r,\epsilon)\simeq c_5 r^{-2} 
\sin\left[2q_{-1}^F r\right]$, where $c_3$ and $c_4$ depend strongly on 
$\epsilon_F$, while $c_5$ is nearly independent of $\epsilon_F$.

\begin{figure}[htb]
\includegraphics[width=0.35\textwidth]{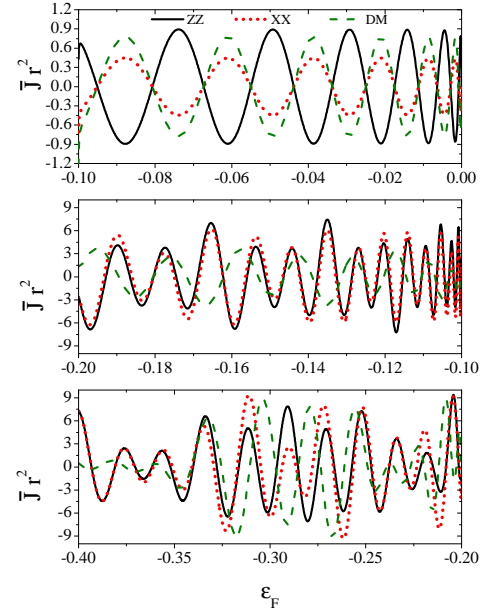}
\caption{(Color online) Comparison of the different components of the RKKY
interaction for different Fermi energy regimes: (a)  $-\epsilon_\Gamma<\epsilon_F<0$, 
(b) $-2\lambda<\epsilon_F<-\epsilon_\Gamma$,
and (c) $\epsilon_F<-2\lambda$. The interimpurity distance is fixed along the zigzag
direction at $r=50$.
Notice different vertical scales.}
\label{fig:R50_EF_m0d1_0_all}
\end{figure}

\noindent
\emph{Fixed distance.}--- 
We now analyze the case where the two impurities remain at a fixed distance along the zigzag direction, 
and analyze the RKKY interaction over a large Fermi energy range. We set $r=50$ in the data shown below.
For $-\epsilon_\Gamma <\epsilon_F<0$  the $\Gamma$ valley does not 
contribute to scattering [Fig.\ \ref{fig:R50_EF_m0d1_0_all}(a)]; all three components 
have similar amplitudes, with XX and DM oscillating in phase with 
each other, but out of phase with ZZ. This indicates an alternation between 
FM (AFM) in plane order and AFM (FM) out-of-plane order as the energy is 
shifted.
When the Fermi energy is positioned in the region 
$-2\lambda<\epsilon_F<-\epsilon_\Gamma$, Fig.\ \ref{fig:R50_EF_m0d1_0_all}(b), 
the ZZ and XX interactions become in phase, while the DM modulation retains a 
longer period. This is caused by the absence of the term 
$I_{G_\Gamma;G_\Gamma}$, because the $\Gamma$ valley is unaffected by the 
spin-orbit interaction.
In this case an isotropic exchange exists at particular values of $\epsilon_F$ 
for a vanishing DM component. 
At deeper Fermi energy, $\epsilon_F<-2\lambda$, with all valleys contributing, 
one finds very interesting behavior:   For 
$\epsilon_F\lesssim -0.35$, there exists another isotropic 
interaction regime with the ZZ and XX components 
contributing equally and the DM term weaker or even zero.

\noindent
\emph{Conclusions}.---
We have shown that inclusion of the $\Gamma$ valley, neglected in previous 
studies, changes predicted magnetic order for RKKY interacting impurities 
deposited on TMD materials. By judicious choice of impurity separation, 
level doping or gating, it is possible to alternate between isotropic and 
anisotropic order as well as to have well defined (or not) in-plane order by 
manipulating the strength of the DM interaction. The results described above 
show behavior that can be readily tested by experiments, such as 
spin polarized STM \cite{Zhou2010, Khajetoorians2012}. Note that although we have 
focused on MoS$_2$, our results are applicable to other dichalcogenides, 
specially WS$_2$ that appears to be easier to dope (or gate).  Characterization of the interaction 
between magnetic impurities with doping level would also provide 
an interesting but direct approach to determine the splitting of the $\Gamma$ valley in real 
systems.  

\emph{Acknowledgments}.---
This work was supported in part by NSF MWN/CIAM grant 
DMR-1108285.

\bibliography{refs_mos2}
\bibliographystyle{apsrev4-1}

\newpage
\onecolumngrid

\section*{Supplemental Material}

\subsection*{Detailed calculation of the RKKY interaction}

We start with the Green's functions in momentum space. For the $K_\tau$ 
valleys, one gets \cite{Parhizgar2013}
\begin{equation}
G_{\tau,s}(q,\epsilon^+) = (\epsilon^+ - \xi)[(\epsilon^+ -
 \xi) (\epsilon^+ + \lambda [1-\tau s])-a^2 q^2]^{-1},
\end{equation}
while at the $\Gamma$ point we have 
\begin{equation}
G_\Gamma(k,\epsilon^+) = 
[\epsilon^+ - E_\Gamma(k)]^{-1}.
\end{equation}
We then apply Fourier transforms as
\begin{equation}\label{eq:FT1}
G_\Gamma(\bo R,\epsilon^+) = \frac{1}{\hat\Omega}\int d\bo k\,\,
 e^{i \bo k \cdot \bo R} 
 G_\Gamma(k,\epsilon^+),
 \end{equation}
and 
\begin{equation}\label{eq:FT2}
G_{\tau,s}(\bo R,\epsilon^+) = \frac{1}{\hat\Omega}\int d\bo k\,\,
 e^{i \bo k \cdot \bo R}  G_{\tau,s}(q,\epsilon^+)= e^{i \bo K_\tau\cdot \bo 
R} g_{\tau,s}(\bo R,\epsilon^+),
 \end{equation}
where 
\begin{equation}
g_{\tau,s}(\bo R,\epsilon^+) = \frac{1}{\hat\Omega} \int d\bo q\,\,
 e^{i \bo q \cdot \bo R}  G_{\tau,s}(q,\epsilon^+),
\end{equation}
and $\hat\Omega$ is the area of the first Brillouin zone. 
Notice that the factor $e^{i \bo K_\tau\cdot \bo R}$ in Eq.\ \eqref{eq:FT2} 
appears because the original Green's functions for valleys 
$K_\tau$ are expressed in terms of the reduced wave vector $\bo q$, while the 
Fourier transform integrates in the momentum $\bo k$ measured from the 
$\Gamma$ point.
From the expressions above, it is easy to observe that $g_{\tau,s}(-\bo 
R,\epsilon^+) = 
g_{\tau,s}(\bo R,\epsilon^+)$, $G_\Gamma(-\bo R,\epsilon^+) = 
G_\Gamma(\bo R,\epsilon^+)$, and $G_{\tau,s}(-\bo R,\epsilon^+) = 
e^{-i\bo K_\tau\cdot \bo R} g_{\tau,s}(\bo R,\epsilon^+)$.

The Fourier transforms involve exponential factors of the form $e^{i \bo k\cdot 
\bo R} = e^{i k R \cos (\theta_{R}-\theta_k)}$, where $\theta_R$ ($\theta_k$) 
is the angle of the interimpurity distance vector (wave vector) measured from 
the positive $x$ axis. 
Using the Jacobi-Anger expansion \cite{Gradshteyn}
\begin{equation}
e^{i k R \cos (\theta_{R}-\theta_k)}
= J_0(kR) + 2\sum_{n=1}^\infty i^n J_n(kR) \cos[n(\theta_R -\theta_k)],
\end{equation}
where $J_n$ are Bessel functions of the first kind and order $n$, we can write
\begin{equation}\label{eq:G_Gamma}
  G_\Gamma(R,\epsilon^+) 
  = \frac{1}{\hat\Omega} \int\! d\bo k \frac{e^{i\bo k \cdot \bo R}}{\epsilon^+ 
- E_{\Gamma}(k)}
  = \frac{2\pi}{\hat\Omega} \int_0^{\infty}\!\! dk \frac{k J_0(k R)}{\epsilon^+ 
- E_{\Gamma}(k)} = -\frac{4\pi \meff t}{\hat\Omega\hbar^2} \int_0^{\infty} 
dk   \frac{k J_0(k R)}{k^2 + \left[i\sqrt{\frac{2 \meff t}{\hbar^2}(\epsilon^+ 
+ \epsilon_\Gamma)} \right]^2}.
\end{equation}
Notice that, after the integration over the angle $\theta_k$, the remaining 
integral over the magnitude of the momentum is evaluated from $0$ to $\infty$. 
To be completely accurate, one should introduce a high momentum cutoff. 
However, 
as one is usually interested in the large distance behavior of the interaction, 
it it expected that the momenta above this cutoff have a negligible 
contribution 
to the integral, so the integration up to $k\rightarrow \infty$ is exact for 
practical purposes. Using the fact that 
\begin{equation}
 \int_0^{\infty} dk \frac{k J_0(k R)}{k^2 + \alpha^2} = K_0[\alpha \sgn (\R 
\alpha) R], 
\end{equation}
where $K_0$ is an order zero modified Bessel function of the second kind, and 
sgn is the sign function, one can rewrite Eq.\ \eqref{eq:G_Gamma} as
\begin{align}
 G_{\Gamma}(R,\epsilon^+) &= -\frac{4\pi \meff t}{\hat \Omega\hbar^2} 
 K_0\bigg[-i\sqrt{\frac{2 \meff t}{\hbar^2}(\epsilon^+ + \epsilon_\Gamma)} 
R\bigg].
\end{align}
At this point it is convenient to define dimensionless parameters: $r\equiv 
R/a$, where $a$ is the closest Mo-Mo distance, $\Omega \equiv 
\hat{\Omega} a^2$, $\gamma \equiv -2 m_{\text{eff}} t a^2/\hbar^2$.
For MoS$_2$, $\meff \simeq -2.6\ m_{\text{el}}$ \cite{Jin2013,Kormanyos_pc}, so 
we get $\gamma \simeq 7.67$. 
Using these conventions, we get the dimensionless Green's function,
\begin{align}
G_{\Gamma}(r,\epsilon^+) &= \frac{2\pi\gamma}{\Omega} 
K_0\bigl[\sqrt{\gamma(\epsilon^+ + 
\epsilon_\Gamma)} r\bigr].
\end{align}
Now we can expand the argument of the Bessel function as
\begin{equation}
 \sqrt{\gamma(\epsilon^+ + \epsilon_\Gamma)} = 
 \begin{cases}
  \sqrt{\gamma(\epsilon+\epsilon_\Gamma)}
  +i\frac{\eta}{2}\sqrt{\frac{\gamma}{\epsilon+\epsilon_\Gamma}} & \epsilon > 
-\epsilon_\Gamma,\\
  \frac{\eta}{2}\sqrt{-\frac{\gamma}{\epsilon+\epsilon_\Gamma}}+
  i\sqrt{-\gamma(\epsilon+\epsilon_\Gamma)} & \epsilon < 
-\epsilon_\Gamma,
 \end{cases}
\end{equation}
such that, for $\eta\rightarrow 0^+$, one gets
\begin{equation}
\begin{split}
 G_{\Gamma}(r,\epsilon^+) &= \frac{2\pi\gamma}{\Omega} 
\Bigl(K_0\bigl[\sqrt{\gamma(\epsilon+\epsilon_\Gamma)}r\bigr]
\theta(\epsilon+\epsilon_\Gamma)
 + K_0\bigl[i\sqrt{-\gamma(\epsilon+\epsilon_\Gamma)}r\bigr]
 \bigl[1-\theta(\epsilon+\epsilon_\Gamma)\bigr]\Bigr),
 \end{split}
\end{equation}
where $\theta$ stands for the Heaviside step function. From this expression, 
one 
can see that the Green's function comprises two parts. The first one, when 
$\epsilon>-\epsilon_\Gamma$, is decaying and accounts for virtual processes in 
which an electron tunnels out of the band. The second one, for 
$\epsilon<-\epsilon_\Gamma$, is oscillating.
We can rewrite this expression by using the identities \cite{Gradshteyn}
\begin{align}
 K_\nu(z) &=  -\frac{i\pi}{2} e^{-i\pi\nu/2} H_\nu^{(2)}(-i z),\\
 H_\nu^{(2)}(z) & = J_\nu(z) - i Y_\nu(z),
 \end{align}
where $Y_\nu(z)$ are Bessel functions of the second kind, and $H_\nu^{(2)}(z)$ 
are Hankel functions. We arrive to
\begin{equation}
 G_{\Gamma}(r,\epsilon^+) = \frac{\pi^2\gamma}{\Omega} 
\Bigl[\frac{2}{\pi} K_0\left[\sqrt{\gamma(\epsilon+\epsilon_\Gamma)}r\right]
\theta(\epsilon+\epsilon_\Gamma)
 -\left(i J_0\big[\sqrt{-\gamma(\epsilon+\epsilon_\Gamma)}r\big] + 
Y_0\left[\sqrt{-\gamma(\epsilon+\epsilon_\Gamma)}r\right]\right)
\left(1-\theta(\epsilon+\epsilon_\Gamma)\right)\Bigr].
\end{equation}
The same procedure can be applied at the $K_\tau$ points
\begin{equation}
\begin{split}
 g_{\tau,s}(r,\epsilon^+) &= -\frac{2\pi}{\Omega} 
(\epsilon^+-\xi)
 \int_0^{\infty}\!\! dq \frac{q J_0(qr)}{q^2 + 
\Bigl(\frac{i}{a}\sqrt{(\epsilon^+-\xi)(\epsilon^+ +\lambda[1-\tau 
s])}\Bigr)^2},
\end{split}
\end{equation}
or
\begin{equation}
 g_{\tau,s}(r,\epsilon^+) = -\frac{2\pi}{\Omega} 
K_0\Bigl[i\sqrt{(\epsilon^+-\xi) (\epsilon^++\lambda[1-\tau 
s])}
\sgn \R \big\{i\sqrt{(\epsilon^+ -\xi)
(\epsilon^+ +\lambda[1-\tau s])}\big\} r\Bigr].
\end{equation}
To get more insight into this expression, let us define $\mathcal{Z}
= (\tilde{\epsilon}^+-\Delta+\lambda)(\tilde\epsilon^++\lambda[1-\tau s])$, so 
$\R \mathcal{Z} 
=\tilde\epsilon^2-\eta^2+\tilde\epsilon(\lambda[2-\tau s]-\Delta) + 
\lambda(\lambda-\Delta)(1-\tau s)$, and
$\I \mathcal{Z} = \eta(2\tilde\epsilon + \lambda(2-\tau s)-\Delta)$.
We have that, for $\eta\rightarrow 0$, $\R \mathcal{Z}=0$ when 
$\tilde\epsilon^2 + \tilde\epsilon(\lambda[2-\tau s]-\Delta) + 
\lambda(\lambda-\Delta)(1-\tau s)=0$, whose solutions are
\begin{equation}
\begin{split}
\tilde\epsilon_{\tau,s}^\pm &= \frac{1}{2} \bigg[\Delta - \lambda(2-\tau s)
\pm \sqrt{[\Delta -\lambda(2-\tau s)]^2 -4\lambda(\lambda -\Delta)(1-\tau s)} 
\bigg].
\end{split}
\end{equation}
For the cases in which $\tau s = 1$, we have $\tilde\epsilon_{1,\ua}^\pm = 
\tilde\epsilon_{-\!1,\da}^\pm \equiv \tilde\epsilon_1^\pm = 
\frac{1}{2} \big[\Delta - \lambda \pm |\Delta-\lambda|\big]$.
Moreover, $(\R \mathcal{Z})' = 2>0$, so we have a parabola with positive 
concavity crossing the $\tilde\epsilon$-axis at 0 and at $\Delta - \lambda>0$.
If we consider that the Fermi energy is always in the valence bands, 
$\epsilon_F<0$, then $\R \mathcal{Z}>0, \forall \epsilon$ if $\tau s=1$.

For the case in which $\tau s=-1$, we have $\tilde\epsilon_{-\!1,\ua}^\pm = 
\tilde\epsilon_{1,\da}^\pm \equiv \tilde\epsilon_{-\!1}^\pm = 
\frac{1}{2} \Bigl[\Delta - 3\lambda \pm (\Delta + \lambda) \Bigr]$, or
\begin{equation}
 \begin{cases}
 \tilde\epsilon_{-\!1}^+ &= (\Delta - \lambda)>0,\\
 \tilde\epsilon_{-\!1}^- &= -2\lambda<0.
 \end{cases}
\end{equation}
Then $\R \mathcal{Z} >0$ if $\epsilon<\tilde\epsilon_{-\!1}^-$ and $\R 
\mathcal{Z} <0$ if $\tilde\epsilon_{-\!1}^-<\epsilon<0$. Finally, $\I 
\mathcal{Z} <0 \,\,\forall\,\, \tau s$.
%
The two independent Green's function are
\begin{equation}
 \begin{split}
 g_{1,\ua}(r,\epsilon^+) &= -\frac{2\pi}{\Omega}(\epsilon^+-\xi) 
 K_0\left[i\sqrt{\epsilon(\epsilon-\xi)}r\right] 
 = \frac{\pi^2}{\Omega}(\epsilon^+-\xi) 
 \Bigl(iJ_0\left[\sqrt{\epsilon(\epsilon-\xi)}r\right] 
+Y_0\left[\sqrt{\epsilon(\epsilon-\xi)}r\right]\Bigr),\\
 g_{-\!1,\ua}(r,\epsilon^+) &= -\frac{2\pi}{\Omega}(\epsilon^+ -\xi)
 \Bigl\{K_0\left[i\sqrt{(\epsilon+2\lambda)(\epsilon-\xi)}r\right] 
 \left[1-\Theta(\epsilon+2\lambda)\right]+ 
K_0\left[\sqrt{-(\epsilon+2\lambda)(\epsilon-\xi)}r\right]
 \Theta(\epsilon + 2\lambda)\Bigr\}\\
 &= \frac{\pi^2}{\Omega}(\epsilon^+-\Delta+\lambda)
\Bigl\{\Bigl(iJ_0\left[\sqrt{(\epsilon+2\lambda)(\epsilon-\xi)}
r\right] 
 + Y_0\left[\sqrt{(\epsilon+2\lambda)(\epsilon-\xi)}r\right]\Bigr) 
\left[1-\Theta(\epsilon + 2\lambda)\right]\\& -\frac{2}{\pi} 
K_0\left[\sqrt{-(\epsilon+2\lambda)(\epsilon-\xi)}r\right]
 \Theta(\epsilon + 2\lambda)\Bigr\}.
 \end{split}
\end{equation}

\begin{figure}[htb]
\includegraphics[width=0.8\textwidth]{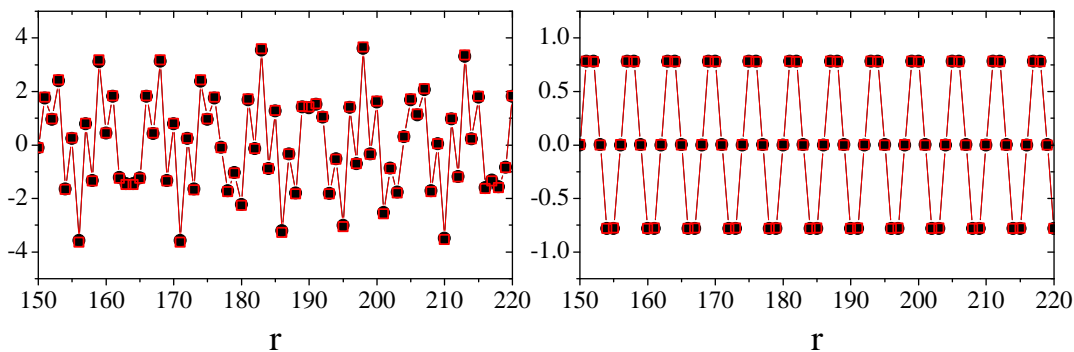}
\caption{Sinusoidal fittings (red squares) to the energy integrals (black 
circles) of $I_{G_\Gamma;g_{1,\ua}}$ (left panel) and $I_{g_{1,\ua};g_{1,\ua}}$ 
(right panel), as discussed in the text.}
\label{fig:fit}
\end{figure}

Now we get the expressions for $\I A_{\alpha,\beta}(\bo r)$, which can be 
subdivided into different terms. To simplify the expressions, we define 
dimensionless wave vectors for the different bands, as a function of the energy
\begin{align}
 k_\Gamma(\epsilon) &= \sqrt{-\gamma(\epsilon+\epsilon_\Gamma)},\nonumber\\
 q_1(\epsilon) &= \sqrt{\epsilon(\epsilon-\xi)},\\
 q_{-1}(\epsilon) &= 
\sqrt{(\epsilon+2\lambda)(\epsilon-\xi)}.\nonumber
\end{align}
We have different components in $A_{\alpha,\beta}(\bo r)$, whose in-plane 
contributions to the susceptibility are given by
\begin{equation}\label{eq:Azz_expanded_sup}
\begin{split}
\I\: A_{z,z}(\bo R)&= 2\Bigl(I_{G_\Gamma;G_\Gamma} + \left[\cos(\bo K_1 \cdot 
\bo R) + \cos(\bo K_{-1} \cdot \bo R)\right]
 \left(I_{G_\Gamma; g_{1,\ua}} + I_{G_\Gamma; g_{-1,\ua}}\right)\\& + 
I_{g_{1,\ua};g_{1,\ua}} + I_{g_{-1,\ua};g_{-1,\ua}}
+ 2\cos\left[(\bo K_{1}-\bo K_{-1}) \cdot \bo R\right] I_{g_{1,\ua}; 
g_{-1,\ua}}\Bigr),
\end{split}
\end{equation}
\begin{equation}\label{eq:Axx_expanded}
\begin{split}
  \I\: A_{x,x}(\bo R) 
 &= 2\Bigl(I_{G_\Gamma;G_\Gamma} + \left[\cos(\bo K_1 \cdot \bo R) + \cos(\bo 
K_{-1} \cdot \bo R)\right] 
\left(I_{G_\Gamma;g_{1,\ua}}+I_{G_\Gamma;g_{-1,\ua}}\right)\\
 &+ \cos\left[(\bo K_{1}-\bo K_{-1}) \cdot \bo R\right] 
 \left[I_{g_{1,\ua};g_{1,\ua}}+ I_{g_{-1,\ua};g_{-1,\ua}}\right]+ 2 
I_{g_{1,\ua};g_{-1,\ua}}\Bigr),
\end{split}
\end{equation}
and
\begin{equation}\label{eq:Axy_expanded}
\begin{split}
 \I\: A_{x,y}(\bo R) &= 2\Bigl(\left[\sin(\bo K_1 \cdot \bo R) - \sin(\bo 
K_{-1} \cdot \bo R)\right] 
\left(I_{G_\Gamma;g_{1,\ua}}-I_{G_\Gamma;g_{-1,\ua}}\right)\\
 &+ \sin\left[(\bo K_{1}-\bo K_{-1}) \cdot \bo R\right]
 \left[I_{g_{1,\ua};g_{1,\ua}}- I_{g_{-1,\ua};g_{-1,\ua}}\right]\Bigr),
\end{split}
\end{equation}
with
\begin{equation}\label{eq:Ggammasq}
 \begin{split}
 I_{G_\Gamma;G_\Gamma}&(r,\epsilon)  = \frac{2\pi^4 \gamma^2}{\Omega^2} 
J_0\left[k_\Gamma(\epsilon)r\right] Y_0\left[k_\Gamma(\epsilon)r\right] 
\left[1-\Theta(\epsilon+\epsilon_\Gamma) \right],
 \end{split}
\end{equation}
\begin{equation}\label{eq:Ggammagp1}
\begin{split}
 I_{G_\Gamma;g_{1,\ua}}(r,\epsilon) &= 
\frac{\pi^4\gamma}{\Omega^2} 
 (\epsilon -\xi)
 \Bigl\{\frac{2}{\pi} J_0\left[q_1(\epsilon)r\right] 
 K_0\left[-ik_\Gamma(\epsilon)r\right]\Theta(\epsilon+\epsilon_\Gamma)\\
&-\Bigl(J_0\left[q_1(\epsilon)r\right] Y_0\left[k_\Gamma(\epsilon)r\right]
+Y_0\left[q_1(\epsilon)r\right] J_0\left[k_\Gamma(\epsilon)r\right]\Bigr)
\left[1-\Theta(\epsilon+\epsilon_\Gamma)\right] \Bigr\},
\end{split}
\end{equation}
\begin{equation}\label{eq:Ggammagm1}
\begin{split}
 I_{G_\Gamma;g_{-1,\ua}}(r,\epsilon) &= \frac{\pi^4\gamma}{\Omega^2} 
 (\epsilon -\xi) \Bigl\{\frac{2}{\pi} J_0\left[k_\Gamma(\epsilon)r\right]
K_0\left[-iq_{-1}(\epsilon)r\right]\left[
1-\Theta(\epsilon+\epsilon_\Gamma)\right]
\Theta(\epsilon+2\lambda)\\
&-\Bigl(J_0\left[k_\Gamma(\epsilon)r\right] Y_0\left[q_{-1}(\epsilon)r\right]
+Y_0\left[k_\Gamma(\epsilon)r\right] J_0\left[q_{-1}(\epsilon)r\right]\Bigr)
\left[1-\Theta(\epsilon+2\lambda)\right] \Bigr\},
\end{split}
\end{equation}
\begin{equation}\label{eq:gp1m1}
\begin{split}
 I_{g_{1,\ua};g_{-1,\ua}}(r,\epsilon) &= \frac{\pi^4}{\Omega^2} 
 (\epsilon -\xi)^2 \Bigl\{\Bigl(J_0\left[q_1(\epsilon)r\right] 
Y_0\left[q_{-1}(\epsilon)r\right]
+Y_0\left[q_1(\epsilon)r\right] J_0\left[q_{-1}(\epsilon)r\right]\Bigr)\\
&\times\left[1-\Theta(\epsilon+2\lambda)\right]-\frac{2}{\pi} 
J_0\left[q_1(\epsilon)r\right] K_0\left[-iq_{-1}(\epsilon)r\right]
\Theta(\epsilon+2\lambda) \Bigr\},
\end{split}
\end{equation}
\begin{equation}\label{eq:gp1sq}
\begin{split}
 I_{g_{1,\ua};g_{1,\ua}}(r,\epsilon)  &= \frac{2\pi^4}{\Omega^2} 
 (\epsilon -\xi)^2 J_0\left[q_1(\epsilon)r\right]
 Y_0\left[q_1(\epsilon)r\right],
\end{split}
\end{equation}
and
\begin{equation}\label{eq:gm1sq}
\begin{split}
 I_{g_{-1,\ua};g_{-1,\ua}}&(r,\epsilon) = \frac{2\pi^4}{\Omega^2}(\epsilon 
-\xi)^2 J_0\left[q_{-1}(\epsilon)r\right] 
Y_0\left[q_{-1}(\epsilon)r\right].
\end{split}
\end{equation}
In order to get the static susceptibility $\chi_{\alpha,\beta}(r)$, we need to 
integrate these expressions over energy. The one corresponding to  Eq.\ 
\eqref{eq:Ggammasq} can be integrated analytically, due to its conventional two 
dimensional parabolic band character \cite{Fischer1975}. Defining $u(\epsilon) 
= 
k_\Gamma(\epsilon) r$, one gets
\begin{equation}
\int_{-\infty}^{\epsilon_F} d\epsilon\,\, I_{G_\Gamma;G_\Gamma}(r,\epsilon) 
= \frac{4\pi^4 \gamma}{\Omega^2 r^2} \int_{u(\epsilon_F)}^\infty du\,\, 
u J_0[u]Y_0[u].
\end{equation}
This integral can be separated into two terms as $\int_{u(\epsilon_F)}^{\infty} 
du = \int_{0}^{\infty} du - \int_0^{u(\epsilon_F)} du$, 
and using the fact that $\int_{0}^{\infty} du= 0$, results in
\begin{equation}
 \begin{split}
\int_{-\infty}^{\epsilon_F} d\epsilon\: I_{G_\Gamma;G_\Gamma}(r,\epsilon)
&= -\frac{4\pi^4 \gamma}{\Omega^2 r^2} \int_0^{u(\epsilon_F)} du\,\, u 
J_0[u]Y_0[u]\\
 &=-\frac{2\pi^4 \gamma[k_\Gamma(\epsilon_F)]^2}{\Omega^2}
\left(J_0[k_\Gamma(\epsilon_F) r] Y_0[k_\Gamma(\epsilon_F) r] + 
J_1[k_\Gamma(\epsilon_F) r] Y_1[k_\Gamma(\epsilon_F) r] \right),
 \end{split}
\end{equation}
For large interimpurity distances, $r \gg 1$, we can approximate the above 
expression as
\begin{equation}
 \int_{-\infty}^{\epsilon_F} d\epsilon\,\, I_{G_\Gamma;G_\Gamma}(r,\epsilon) 
 = \frac{2\pi^3\gamma}{\Omega^2} \frac{\sin\left[2 k_\Gamma(\epsilon_F) 
r\right]}{r^2}.
\end{equation}
The remaining integrals should be performed numerically and the integration 
requires some care. They should be regularized by a smooth energy cutoff 
function, as discussed in Ref.\ \onlinecite{Saremi2007}. We tried different 
cutoff functions in order to test convergence.
As discussed in the main text, these integrals can be fitted by sinusoidal 
functions with prefactors that can depend or not of the Fermi energy. Two 
examples are shown in Fig.\ \ref{fig:fit} for $\epsilon_F = -0.174$. The left 
panel shows the sinusoidal fitting to $r^2 \int_{-\infty}^{\epsilon_F} 
d\epsilon\,\, I_{G_\Gamma;g_{1,\ua}}(r,\epsilon) \simeq 0.24 \sin(1.278\: r)$, 
and the right panel corresponds to the one for $r^2 \int_{-\infty}^{\epsilon_F} 
d\epsilon\,\, I_{g_{1,\ua};g_{1,\ua}}(r,\epsilon) \simeq 0.45 \sin(1.047\: 
r)$.

\subsection*{Angular dependence of the interaction}

The angular dependence enters in different combinations of $\bo K_\tau\cdot \bo 
R$ [see Eqs.\ (\ref{eq:Azz_expanded_sup}) - (\ref{eq:Axy_expanded})].
The vector that connects the impurities can be written as 
$\bo R = m \bo a_1 + n \bo a_2$, where $\bo a_{1/2} = \frac{a}{2}(1,\pm 
\sqrt{3})$, $m,n \in \mathbb{Z}$, and 
$\bo a_i$ are the primitive vectors (see Fig.\ 1 in the main text); 
in dimensionless form, $\bo r = \bo R/a = 
\frac12 \left(m+n,\sqrt{3}(m-n)\right)$, with $r= 
\sqrt{(m-n)^2+mn}$.
We can also define dimensionless valley vectors as $\Kt = a \bo 
K_\tau$, such that $\bo K_\tau \cdot \bo R = \Kt \cdot \bo r = 
\pi\left[\left(\frac{\tau}{3}+1\right)m 
+\left(\frac{\tau}{3}-1\right)n\right]$, 
and $(\Kp - \Km)\cdot \bo r = \frac{2\pi}{3}\left(m+n\right)$.
Three zigzag directions are possible, for $(m,n)$ combinations given by 
$(p,0)$, $(0, p)$ and $(p, p)$, with integer $p$, and $r=|p|$, where the 
angular coefficients are shown in Table \ref{tab:zigzag}.
\begin{table}
\begin{tabular}{|l | l | l |}
  \hline
  \multicolumn{3}{|c|}{Zigzag}\\
  \hline
  Coefficient & Direction & Sequence\\
  \hline
  $\cos[(\Kp -\Km)\cdot \bs r]$ & all & \multirow{3}{*}{$1$, $-\frac12$, 
  $-\frac12$,$\cdots$}\\
  $\cos[\Kp \cdot \bs r]$ & all&\\
  $\cos[\Km \cdot \bs r]$ & all&\\
  \hline
  $\sin[(\Kp - \Km)\cdot \bs r]$ & (p, 0); (0, p) & \multirow{3}{*}{$0$, 
$\frac{\sqrt{3}}{2}$, $-\frac{\sqrt{3}}{2}$,$\cdots$}\\
$\sin[\Kp \cdot \bs r]$ & (p, p) & \\
$\sin[\Km \cdot \bs r]$ & (p, 0); (0, p) & \\
\hline
$\sin[(\Kp -\Km)\cdot \bs r]$ & (p, p) & \multirow{3}{*}{$0$, 
$-\frac{\sqrt{3}}{2}$, 
$\frac{\sqrt{3}}{2}$,$\cdots$}\\
$\sin[\Kp \cdot \bs r]$ & (p, 0); (0, p) & \\
$\sin[\Km \cdot \bs r]$ & (p, p) & \\
\hline
\end{tabular}
\caption{Sequences of angular dependent coefficients for the zigzag directions.}
\label{tab:zigzag}
\end{table}
The armchair directions are given by $(2p,p)$, $(p,2p)$, $(p,-p)$, so 
$r =\sqrt{3}|p|$. The coefficients in the armchair direction are simpler 
than for zigzag, as $\cos(\Kt \cdot \bs r) = \cos[(\Kp - \Km)\cdot \bs r] = 
1$ and $\sin(\Kt \cdot \bs r) = \sin[(\Kp - \Km)\cdot \bs r] = 0$.
This means that in the armchair direction the DM component is always 
zero due to symmetry.

\end{document}